\documentclass[useAMS,usenatbib,usegraphicx]{mn2e}

\usepackage{amsmath,amssymb,amsbsy,amsfonts}
\usepackage{amssymb}
\usepackage[british]{babel}
\usepackage{mdwmath}
\usepackage{cancel}
\usepackage{color}
\usepackage{url}

\title[Gravitational lensing in extended \(f(\chi)=\chi^{3/2}\)
gravity]{Gravitational lensing with \(f(\chi)=\chi^{3/2} \) gravity in
accordance with astrophysical observations}

\author[S. Mendoza, et al.]
       {S. Mendoza$^{1}$\footnotemark[1], T. Bernal$^{1}$\footnotemark[1], 
       X.  Hernandez$^{1}$\footnotemark[1], J.C.  
       Hidalgo$^{1,2}$\footnotemark[1] \& L.A. Torres$^{1}$\thanks{E-mail
       address: sergio@astro.unam.mx~(SM), tbernal@astro.unam.mx~(TB),
       xavier@astro.unam.mx~(XH), ltorres@astro.unam.mx~(LAT)} \\
           $^{1}$Instituto de Astronom\'{\i}a, Universidad Nacional
           Aut\'onoma de M\'exico, AP 70-264, Distrito Federal 04510,
           M\'exico \\
           $^2$Departamento de F\'{\i}sica, Instituto Nacional de
	       Investigaciones Nucleares, La Marquesa Ocoyoacac, 52750,
               M\'exico
       }

\date{\today}
\volume{VER14}
\pagerange{\pageref{firstpage}--\pageref{lastpage}}

\begin{document}

\maketitle
\label{firstpage}

\begin{abstract}
  In this article we perform a second order perturbation analysis of
the gravitational metric theory of gravity \( f(\chi) = \chi^{3/2}
\) developed by \citet{bernal11}.  We show that the theory accounts
in detail for two observational facts: (1) the phenomenology of
flattened rotation curves associated to the Tully-Fisher relation
observed in spiral galaxies, and (2) the details of observations of
gravitational lensing in galaxies and groups of galaxies, without the
need of any dark matter.  We show how all dynamical observations on flat
rotation curves and gravitational lensing can be synthesised in terms
of the empirically required metric coefficients of any metric theory
of gravity.  We construct the corresponding metric components for the
theory presented at second order in perturbation, which are shown to be
perfectly compatible with the empirically derived ones.  It is also shown
that under the theory being presented, in order to obtain a complete
full agreement with the observational results, a specific signature of
Riemann's tensor has to be chosen.  This signature corresponds to the one
most widely used nowadays in relativity theory.  Also, a computational
program, the MEXICAS (Metric EXtended-gravity Incorporated through a
Computer Algebraic System) code, developed for its usage in the Computer
Algebraic System (CAS) Maxima for working out perturbations on a metric
theory of gravity, is presented and made publicly available.
\end{abstract}


\section{Introduction}
\label{introduction}

  When Einstein introduced his theory of general relativity, an
astrophysical prediction for the motion of the planet Mercury (a massive
particle) through its orbit was made \citep{einstein16}.  The second step
was to test general relativity through the deflection of light (massless
particles) coming from stars appearing near the Sun's limb during a solar
eclipse \citep{eddington20}.   Both observations constituted the first coherent
steps towards the solid foundation of general relativity, a theory
capable of describing gravitation through a correct relativistic description.

  In this sense, any metric theory of gravity must be compatible with
both kinds of observations, the dynamical ones for massive particles
and the observations of the deflection of light for massless particles.
The correct approach is extensively described in the monograph
by \citet{will93} where it is shown that when working with the weak field limit
of a relativistic theory of gravity in a static spherically spacetime,
the dynamics of massive particles determine the functional form of the
time component of the metric, while the deflection of light determines the
form of the radial one \citep[see also][and references therein]{will06}.

  To order of magnitude and through a first perturbation analysis,
\citet{bernal11} have shown that it is possible to recover flat
rotation curves and the Tully-Fisher relation (i.e. a  MONDian-like
weak field limit) from a metric theory of gravity, which includes the
mass of the system in the gravitational field's action. Such limit
is of high astrophysical relevance at the scales of galaxies, where
MOND accurately describes the rotation curves of spiral galaxies and
the Tully-Fisher relation without the need of dark matter \citep[see
e.g.][]{milgrom83b,famaey11}.  In this article we show the strength
of the calculations made by \citet{bernal11} by doing an extensive
analysis from perturbation theory for a static spherically symmetric
metric and show that in the weak field limit our results are in perfect
agreement not only with the Tully-Fisher relation, but are also in
exact accordance with observations of gravitational lensing over a wide
range of astrophysical scales.

  Extensions to Einstein's general theory of relativity have been
proposed since the publication of the theory itself \citep[see
e.g.][]{schimming04}.  However, it has not been until recent times that
observations at different mass and length scales have concluded that
in order to keep Einstein's field equations valid, unknown dark matter
and energy entities need to be added to the theory.  In this article,
a complementary approach is taken where the existence of these
unknown dark entities is not required.  We show the theory built
by \citet{bernal11} to be in accordance not only with the very well
established observations of the dynamics of massive particles through the
Tully-Fisher relation, but also with the dynamics of massless particles
through the bending of light as astrophysically observed.

  \citet{mendoza07} and~\citet{yetli-tesis} showed for the first time
that metric theories of gravity are capable of producing more deflection
of light that the one produced by Einstein's general relativity.
This was done using the metric theory of gravity constructed by
\citet{sobouti07}.  The implications of this result invalidated the so
called no-go theorem for metric \( f(R) \) theories of gravity proposed
by \citet{soussa03a,soussa03b}. Furthermore, in the present work we show
that it is possible to explain the observed gravitational lensing for
galaxies, and groups of galaxies without the need of invoking dark matter.
Developments by  \citet[see e.g.][]{capozziello06,horv12,nzioki11} on weak
and strong lensing regimes of extended metric theories of gravity have
followed the work by \citet{mendoza07} but are not of general validity
with respect to different astrophysical observations.

 Testing any metric theory of gravity against observations can be
cumbersome. From an action principle one must derive field equations,
which in principle, have to be solved for e.g. in spherically symmetric
spacetimes. The solutions to this lead to metric coefficients which in
turn, with the use of the geodesic equation, yield orbits for massive
and massless particles, to be then compared against astrophysical
observations. These last are varied and diverse e.g., centrifugal
equilibrium orbits at a variety of radii, for systems having total masses
spanning several orders of magnitude, and the observed shears and caustic
positions of gravitational lensing observations.

Fortunately, we have derived a much more direct and generic approach.
First, dynamical observations regarding the amplitudes of galactic flat
rotation curves satisfy a well known scaling with the fourth root of the
total baryonic content: the Tully-Fisher relation.  To second order in
perturbations of the velocity measured in units of the speed of light,
this can be shown to imply a definite  empirical prescription for the time
component of any metric theory not requiring dark matter.  Second, we
show that all gravitational lensing observations on elliptical and spiral
galaxies, as well as for groups of galaxies can be synthesised as the
requirement for the same isothermal total matter distribution as needed to
explain the observed spiral rotation curves and dynamics about elliptical
galaxies, if one assumes Einstein's general relativity.  
From studying directly the lens equation in general relativity, this
implies a bending angle which is independent of the impact parameter,
and which scales with the square root of the total baryonic mass of a
system. It can then be shown that this, in combination with the empirical
time component of the metric mentioned above, leads to a definitive
empirical form for the radial component, for any metric theory not
requiring dark matter.

  Thus, we synthesise all dynamical and gravitational lensing
astrophysical observations at galactic and galaxy group scales, into
empirical time and radial metric components of a spherically symmetric
metric given at second order in perturbation. It is through comparing
the above to perturbed metric coefficients to the same order coming
from the metric theory treated in this paper that we are able to show
its full compatibility with all relevant dynamical and gravitational
lensing astrophysical observations.

  The article is organised as follows.  In section~\ref{weak-field-limit}
the concept of weak field limit for a static spherically symmetric metric
is established and we define the relevant orders of perturbation to be
used throughout the article.  In section~\ref{field-equations} we perturbe
the vacuum field equations for the metric theory built by \citet{bernal11}
and show that for a point mass source they closely resemble the ones
usually adopted in \( f(R) \) gravity in vacuum. However, these equations
slightly differ under the approximations of the mass and length scales
associated to galaxies and groups of galaxies -where gravity is expected
to differ from Einstein's general relativity in the absence of any dark
matter component.  In section~\ref{lowest-order} we obtain the solution
for the Ricci scalar up to the second order from the perturbed field
equations and discuss the importance of the signature in the Riemann
tensor to yield the correct results.  In section~\ref{metric-components}
we obtain the coefficients of the metric up to the second order
in perturbation.  In section~\ref{metric-coefficients} we obtain
the metric coefficients up to the second order in an empirical way,
without reference to any specific metric theory of gravity, using the
dynamical phenomenology of galaxies and groups of galaxies and the
gravitational lensing produced by these objects.  In that section  we
also compare the metric coefficients obtained in~\ref{metric-components}
with those empirically obtained and show full consistency.  Finally in
section~\ref{discussion} we discuss our results.

\section{The weak field limit}
\label{weak-field-limit}

  An excellent account of perturbation theory applied to metric theories
of gravity (in particular general relativity) can be found in the
monograph written by \citet{will93}. More recently, \citet{stabile09}
have developed a perturbation analysis technique useful when dealing with
lenses in \( f(R) \) gravity.  In this section we define the relevant
properties of the perturbation theory having in mind applications to
the metric theory developed by \citet{bernal11}.

  Let us consider a fixed point mass \( M \) at the centre of coordinates
generating a gravitational field.  Under these considerations, the
spacetime is static and its spherically symmetric metric \( g_{\mu\nu} \) is
generated by the interval

\begin{equation}
  \mathrm{d}s^2 = g_{\mu\nu} \mathrm{d}x^\mu \mathrm{d} x^\nu  
    = g_{00} \, c^2 \mathrm{d}t^2 + g_{11}\mathrm{d}r^2
	-r^2 \mathrm{d} \Omega^2.
\label{metric}
\end{equation}

\noindent  In the previous equation and in what follows, Einstein's
summation convention over repeated indices is used. Greek indices take
values \(0,1,2,3\) and Latin ones \(1,2,3\). As such, in spherical
coordinates  \( (x^0,x^1,x^2,x^3) = (ct,r,\theta,\varphi) \), where \(
c \) is the speed of light, \( t \) is the time coordinate, \( r \) the
radial one, and \( \theta \) and \( \varphi \) are the polar and azimuthal
angles respectively. Also, the angular displacement \( \mathrm{d}
\Omega^2 := \mathrm{d}\theta^2 + \sin^2\theta \, \mathrm{d}\varphi^2 \).
Due to the symmetry of the problem, the unknown functions \( g_{00} \)
and \( g_{11} \) are functions of the radial coordinate \( r \) only.
Note also that We choose a \( (+,-,-,-) \) signature for the the spacetime
metric, which we maintain throughout the article.

  The radial component of the geodesic equations

\begin{equation}
  \frac{ \mathrm{d}^2 x^\alpha }{ \mathrm{d} s^2 } +
\Gamma^\alpha_{\beta\lambda} \frac{
    \mathrm{d} x^\beta}{ \mathrm{d} s } \frac{ \mathrm{d} x^\lambda }{ 
    \mathrm{d} s } = 0 \, ,
\label{geodesic}
\end{equation}

\noindent for the metric~\eqref{metric} in the weak field limit, i.e.
when the speed of light \( c \to \infty \), is given by

\begin{equation}
  \frac{1}{c^2} \frac{\text{d}^2 r}{\text{d} t^2} =
   \frac{1}{2} g^{11} g_{00,r} ,
\label{radial:geodesic}
\end{equation}

\noindent where the subscript \( ( \ )_{,r} := \partial / \partial r
\) denotes the derivative with respect to the radial coordinate. In
the above relation we have assumed that for the weak field limit \(
\text{d}s = c \ \text{d}t \) and since the velocity \(v \ll c\) then \(
v^i  \ll \text{d}x^0 / \text{d}t\) with \( v^i := \left( \mathrm{d}r /
\mathrm{d}t,\ r \mathrm{d} \theta / \mathrm{d}t,\ r\sin \theta \mathrm{d}
\varphi / \mathrm{d}t \right) \). In the strong \(c \rightarrow \infty
\) limit, both sides of the above equation vanish simultaneously. Thus,
the condition of the right hand side of equation~\eqref{radial:geodesic}
vanishing in the \(v \ll c\) provides a consistency check on the results
of the following sections, where a perturbative solution to the metric
is developed.

  In this limit, a particle bound to a circular orbit about the mass \(
M \) experiences a centrifugal radial acceleration given by:

\begin{equation}
  \frac{\text{d}^2 r}{\text{d} t^2} =  \frac{v^2}{r},
\label{centrifugal}
\end{equation}

\noindent for a circular or tangential velocity \( v \). The preceding
equation is a kinematical relation of general validity 
and does not introduce any particular assumption of the gravitational
theory.

  When material particles are used as test particles in the weakest 
limit of the theory, the metric takes the form 
\citep[see e.g.][]{daufields}:

\begin{align}
  g_{00} &= 1 + \frac{ 2 \phi }{ c^2 },  &\quad g_{11} &= -1, \nonumber \\
  g_{22} &= -r^2, &\quad g_{33} &= -r^2 \sin^2{\theta},
\label{weakest-metric}
\end{align}

\noindent for a Newtonian gravitational potential \( \phi \).  The above
equations are only used when analysing the motion of material particles
when gravity is very weak \citep[see e.g.][]{will93}.
In order to demand greater accuracies of the theory and to 
recover exact results for the motion of massless particles, i.e.
to accurately describe the bending of light rays, the following term in 
the expansion of $g_{11}$ must also be considered.

  For a particle on circular motion about the mass \( M \) in the
weak field limit, the lowest order of the theory is obtained when the
left-hand side of equation~\eqref{radial:geodesic} is of order \( v^2
/ c^2 \) and when the right-hand side is of order \( \phi / c^2 \).
Both are just orders \( \mathcal{O}(1/c^2) \) of the theory, or simply
\( \mathcal{O}(2) \). As such, when lower or higher order corrections of the
theory are introduced we will use the notation \( \mathcal{O}(n) \)
for \( n=0,1,2,\ldots\) meaning \( \mathcal{O}(0), \ \mathcal{O}(c^{-1}), \
\mathcal{O}(c^{-2}),\ldots \) respectively.

  Having in mind further astrophysical applications (of motion of material
particles and bending of light -massless particles), we expand the
metric \( g_{\mu\nu} \) about a flat Minkowski metric \( \eta_{\mu\nu}
:= \text{diag}(1,-1,-1,-1) \) up to the second order in time and radial
position in such a way that

\begin{equation}
\begin{split}
	g_{00} =& g_{00}^{(0)} + {}g_{00}^{(2)} = 1 + {}g_{00}^{(2)}
	  + \mathcal{O}(4),  \\
	g_{11} =& {}g_{11}^{(0)} + {}g_{11}^{(2)} = - 1 +
	  {}g_{11}^{(2)} + \mathcal{O}(4),  \\
	g_{22} =& {}g_{22}^{(0)} = -r^2,  \\
	g_{33} =& g_{22} \sin^2 \theta,
\end{split}
\label{metric-perturbed}
\end{equation}

\noindent where the superscript \( (p) \) denotes the order \( \mathcal{O}(p)
\) at which a particular quantity is approximated.
From equations~\eqref{metric-perturbed} it follows that the contravariant
metric components are given by

\begin{equation}
\begin{split}
	g^{00} =& g^{00(0)} + g^{00(2)} = 1 - {}g_{00}^{(2)} +
	  \mathcal{O}(4),  \\
	g^{11} =& g^{11(0)} + g^{11(2)} = - 1 - g_{11}^{(2)} +
	\mathcal{O}(4),  \\
	g^{22} =& g^{22(0)} = - 1/r^2, 	 \\
	g^{33} =& g^{22} / \sin^2 \theta.
\end{split}
\label{metric:exp-up}
\end{equation}

\noindent  In fact, to the lowest order of perturbation, we need
to find the time \( g_{00}^{(2)} \) and radial \( g_{11}^{(2)}
\) metric components up to the second order to compare with the
astrophysical observations of material particles and bending of light
\citep{will93,will06}.  Note that in keeping with the assumption
of spherical symmetry for the matter configurations to be studied,
we consider no perturbations on the angular terms of the metric. This
assumption thus limits the applicability of all our following results to
systems not far from spherical symmetry, e.g. the spheroidal elliptical
galaxies about which gravitational lenses are often detected.

\section{Field equations}
\label{field-equations}

  For the case of a point-mass source generating a gravitational field,
\citet{bernal11} have proposed an extended gravitational 
field's action in the metric approach given by:

\begin{equation}
   S_\text{f}  = - \frac{ c^3 }{ 16 \pi G L_M^2 } \int{ f(\chi) \sqrt{-g}
     \, \mathrm{d}^4x} \, ,
\label{action}
\end{equation}

\noindent for any arbitrary dimensionless function \( f(\chi) \) of 
the dimensionless Ricci scalar:

\begin{equation}
  \chi := L_M^2 R,
\label{chi}
\end{equation}

\noindent where \( R \) is the standard Ricci scalar and

\begin{equation}
	L_M = \zeta r_\textrm{g}^{1/2} l_M^{1/2},  
\label{def:LM}
\end{equation}

\noindent is a length scale with

\begin{equation}
	r_\text{g} := \frac{GM}{c^2}, \qquad l_M:=\left( \frac{GM}{a_0}
	\right)^{1/2},
\label{defs}
\end{equation}

\noindent with \(l_M\) the mass-length scale of the system defined
by \cite{mendoza11},  \(a_0 := 1.2 \times 10^{-10} \, \mathrm{m} /
\mathrm{s}^2 \) is Milgrom's acceleration constant \citep[see e.g.][and
references therein]{famaey11} and \(\zeta\) is a coupling constant of
order one which has to be calibrated through astrophysical observations.
This \( f(\chi) \) theory was constructed through the inclusion of
\(a_0\) as a fundamental physical constant, which has been shown to be of
astrophysical and cosmological relevance \citep[see e.g.][]{bernal11,
bernal11b,carranza12,mendoza11,mendoza12,hernandez10,hernandez12a,
hernandez12b}.  Equation~\eqref{action} is understood as a particular
case of a fuller formulation where the details of the mass distribution
appear inside of the action integral, in such a way that for a fixed
point mass, the result is the action~\eqref{action}, as will be more
fully discussed towards the end of this section.

  Following the description of \citet{bernal11} the matter action takes
its ordinary form:

\begin{equation}
	S_\text{m} = - \frac{ 1 }{ 2 c } \int{ {\cal L}_\text{m} \, \sqrt{-g} \,
    \mathrm{d}^4x } \, ,
\label{matter-action}
\end{equation}

\noindent with \( {\cal L}_\text{m} \) the Lagrangian matter density of the
system. The null variation of the complete action, i.e. \( \delta
\left( S_\text{f} + S_\text{m} \right) = 0 \), with respect to the metric
\( g_{\mu\nu} \) yields the following field equations:

\begin{equation}
  \begin{split}
  f^\prime(\chi) \chi_{\mu\nu}   -  \frac{1}{2} f(\chi)
     g_{\mu\nu}  &-  L_M^2 \left( \nabla_\mu \nabla_\nu -g_{\mu\nu}
   \Delta \right) f'(\chi) 
     \\
   & = \frac{ 8 \pi G L_M^2 }{ c^4} T_{\mu\nu},
  \end{split}
\label{field:eqs}
\end{equation}

\noindent where the Laplace-Beltrami operator has been written as
\(\Delta := \nabla^\mu \nabla_\mu\), the prime denotes the derivative
with respect to the argument and the energy-momentum tensor \( T_{\mu\nu}
\) is defined through the standard relation \( \delta S_{\textrm m}
= - (1/2c) T_{\alpha\beta} \delta g^{\alpha\beta} \).  Also, in
equation~\eqref{field:eqs}, the dimensionless Ricci tensor is defined as:

\begin{equation}
	\chi_{\mu\nu} := L_M^2 R_{\mu\nu} \, ,
\label{def:chi}
\end{equation}

\noindent where  \(R_{\mu\nu}\) is the standard Ricci tensor.

  The trace of equations \eqref{field:eqs} is

\begin{equation}
  f^\prime(\chi) \, \chi  - 2 f(\chi) + 3 L_M^2  \, \Delta  f^\prime(\chi) =
    \frac{ 8 \pi G L_M^2 }{ c^4} T \, ,
\label{trace}
\end{equation}

\noindent where \( T := T^\alpha_\alpha \).

    To order of magnitude approximation, where 
\( \mathrm{d} / \mathrm{d} \chi \approx 1 / \chi \), \(
\Delta \approx - 1 / r^2  \) and the mass density \( \rho \approx  M /
r^3 \), \citet{bernal11} have shown that the trace~\eqref{trace} equation
takes the following form:

\begin{equation}
   \chi^b  \left( b - 2 \right) - 3 b L_M^2  \frac{ \chi^{(b-1)} }{ r^2 }
     \approx \frac{ 8 \pi G M L_M^2 }{ c^2 r^3}.
\label{trace-order}
\end{equation}

\noindent for a power-law form:

\begin{equation}
	f(\chi) = \chi^b .
\label{power-law}
\end{equation}

\noindent As shown by \citet{bernal11}, the third term on the left-hand
side of  equation~\eqref{trace} dominates over the first two when the radius of
curvature \( R_\text{c} \approx R^{-1/2} \) of spacetime is such that \(
R_\text{c} \gg r \) and so, this corresponds to the region where MONDian
effects are expected to appear.

  \citet{bernal11} and \citet{mendoza12} have shown that the function
\( f(\chi) \) must satisfy the following limits:

\begin{equation}
	f(\chi) = 
	\begin{cases}
          \chi, \qquad \text{when } \chi \gg 1, \\
          \chi^{3/2}, \quad \text{when } \chi \ll 1.
        \end{cases}
\label{f-chi-real}
\end{equation}

\noindent The limit \( \chi \gg 1 \) recovers Einstein's general
relativity and the condition \( \chi \ll 1 \) yields a relativistic
version of MOND. In this last regime, the first two terms on the right-hand
side of the trace~\eqref{trace} are smaller than the third and so
\citep[cf.][]{bernal11}: 

\begin{equation}
  f^\prime(\chi) \, \chi  - 2 f(\chi) \ll  3 L_M^2  \, \Delta  f^\prime(\chi),
\end{equation}

\noindent at all orders of approximation, and so the trace~\eqref{trace}
is given by:

\begin{equation}
	3 L_M^2 \Delta f^\prime(\chi) = \frac{8 \pi G L_M^2}{c^4} T .
\label{delta:chi}
\end{equation}

\noindent Since we are interested in the field produced by a point
mass \( M \), then the right-hand side of equations~\eqref{field:eqs}
and~\eqref{delta:chi} are null away from the source and so, the last relation
in vacuum can be rewritten as:

\begin{equation}
	\Delta f^\prime(\chi) = 0 .
\label{null:delta}
\end{equation}

  As shown by \citet{bernal11}, the relation \( f(\chi)= \chi^{3/2} \) yields
the correct MONDian non-relativistic limit.  However, for the sake of
generality we will assume in what follows that the function \( f(\chi)
\) is of power-law form~\ref{power-law}.  In this case, 
relation~\eqref{null:delta} is equivalent to

\begin{equation}
	\Delta f^\prime(R) = 0 ,
\label{null:deltaR}
\end{equation}

\noindent to all orders of approximation for a power-law function of
the Ricci scalar

\begin{equation}
 f(R) = R^b.
\label{eq3b}
\end{equation}

  Substitution of the power-law function~\eqref{power-law} in the
null variations of the gravitational field's  action~\eqref{action} in vacuum
means that

\begin{equation}
  \delta S_\text{f}  = - \frac{ c^3 }{ 16 \pi G } L_M^{2(b-1)} \delta\int{
     R^b \sqrt{-g} \, \mathrm{d}^4x} = 0, 
\label{eq03}
\end{equation}

\noindent and so

\begin{equation}
  \delta\int{ R^b \sqrt{-g} \, \mathrm{d}^4x }   = 0.
\label{eq03a}
\end{equation}

\noindent This equation gives the same field equations as the
null variation of the action for a standard power-law metric \( f(R)
\) theory~\eqref{eq3b} in vacuum. With this in mind, we can follow the
standard perturbation analysis for \( f(R) \) restricted by the constraint
equation~\eqref{null:deltaR} needed to yield the correct MOND-like limit.
Since we are only interested in a power-law description of gravity far
away from general relativity (cf. equation~\eqref{f-chi-real}), then in
what follows we use the standard \( f(R) \) field equations for vacuum
as described by \citet{capozziello-book} for a power-law description
of gravity given by equation~\eqref{eq3b} with \( b = 3/2 \), with the
constraint~\eqref{null:deltaR}. To follow their notation, we write the
field equations~\eqref{field:eqs} in vacuum as

\begin{equation}
	f^{\prime}(R) R_{\mu\nu} - \frac{1}{2} f(R) g_{\mu\nu} +
   \mathcal{H}_{\mu\nu} = 0 \, ,
\label{einstein:eqs}
\end{equation}

\noindent where the fourth-order terms are grouped into the following
quantity:

\begin{equation}
	\mathcal{H}_{\mu\nu} := - \left( \nabla_\mu \nabla_\nu
    - g_{\mu\nu} \Delta \right) f^{\prime}(R) \, .
\end{equation}

\noindent The trace of equation~\eqref{einstein:eqs} is thus given by

\begin{align}
	f^{\prime}(R) R - 2f(R) + \mathcal{H} = 0 \, ,
\label{trace:einstein}
\end{align}

\noindent with

\begin{equation}
	\mathcal{H} :=  \mathcal{H}_{\mu\nu}g^{\mu\nu} = 3 \Delta f^\prime(R) .
\end{equation}

  The mathematical form of the field's action~\eqref{action} includes
the Schwarzschild mass (through \( L_M \)) in the description of the
gravitational field.  This is usually not the case for the description of
the gravitational field since the matter content is generally assumed to
appear only in the the matter action~\eqref{matter-action}.  Following
the remarks by \citet{sobouti07} and \citet{mendoza07}, where similar
conclusions were reached, one should expect extensions to the theory even
at the fundamental level of the action.   For the case of a general matter
distribution it is not evident what path to follow.  As explained by
\citet{carranza12} and \citet{mendoza12}, for systems with a high degree
of symmetry (such as the Friedmann-Lema\^{\i}tre-Robertson-Walker -FLRW-
Universe or a spherically symmetric distribution of matter) the action
may be postulated as:

\begin{equation}
   S_\text{f}  = - \frac{ c^3 }{ 16 \pi G  } \int{ \frac{f(\chi)}{ L_M^2}
   \sqrt{-g} \, \mathrm{d}^4x} \, ,
\label{extended-action}
\end{equation}

\noindent where the mass-energy is given by \citep[see e.g.][]{MTW}

\begin{equation}
   M(r) = 4 \pi \int_0^r{ \rho(r) \, r^2 \mathrm{d} r.}
\label{mass-r}
\end{equation}

\noindent For the case of the FLRW universe, the upper limit of the
previous integral is taken as the Hubble
radius~\citep[cf.][]{carranza12,mendoza12}.  The connection between the
action~\ref{extended-action} and the \( F(R,T) \) theory described by
\citet{harko11} is then evident through the identification

\begin{equation}
  F(R,T) := \frac{ f(\chi) }{ L_M^2 }.
\label{harko-connection}
\end{equation}

\noindent  The field equations then follow through the full formal
variation of the action with respect to both \( R \) and \( T \)
\citep[see e.g.][]{harko11,mendoza12}.

  The general description of the gravitational theory is by no means
complete, and further investigation needs to be carried out in this
direction.  We only mentioned one possible generalisation of the simple
point mass description by \citet{bernal11} for completeness.  In any
case, the lensing phenomena we are interested in occur sufficiently far
away from the matter distribution producing it, that these can be
correctly described as point mass sources.

  In what follows the sign convention used in the definition of
the Riemann tensor becomes a relevant point.  As discussed in
appendix~\ref{appendix1}, the solutions to the differential field
equations of any \( f(R) \) theory of gravity greatly depend on the
signature chosen for Riemann's tensor.   Two different choices of
signature bifurcate on the solution space, a property which does not
appear in Einstein's general relativity.  This is not surprising as it
mirrors the analogous unfolding of the metric and Palatini approaches in
\( f(R) \) gravity, which does not appear in Einstein's \( f(R) = R \)
theory \citep[see e.g.][]{olmo11}.  Throughout the article we select
a particular branch of solutions given by the nowadays almost standard
definition of Riemann's tensor in equation~\eqref{riemanndef}.

  In dealing with some of the cumbersome algebraic manipulations that a
perturbation to an \( f(R) \) theory of gravity presents, we have used
the Computer Algebra System (CAS) Maxima to facilitate the computations.
The MEXICAS (Metric EXtended-gravity Incorporated through a Computer
Algebraic System) code (Copyright of T. Bernal, S. Mendoza and L.A. Torres
and licensed with a GNU Public License Version 3)  we wrote for this
is described in appendix~\ref{appendix2} and can be downloaded from:
http://www.mendozza.org/sergio/mexicas.  Further development on the
treatment of the field equations by the MEXICAS code is described in
appendix~\ref{appendix3}.

  For the case of a static spherically symmetric spacetime~\eqref{metric} it 
follows that

\begin{equation}
\begin{split}
  \mathcal{H}_{\mu\nu} = & - f^{\prime\prime} \bigg\{R_{, \mu\nu} -
    \Gamma^1_{\mu\nu} R_{,r} - g_{\mu\nu} \bigg[ \bigg(
    g^{11}_{~,r} \bigg. \bigg. \bigg.  \\
   & \left. \left. \bigg.  + g^{11} \left( \ln{ \sqrt{-g} }
    \right)_{,r} \right) R_{,r} + g^{11} R_{,rr} \right] \bigg\} \\
   & - f^{\prime\prime\prime} \bigg\{ R_{,\mu}R_{,\nu} -
    g_{\mu\nu}g^{11} R_{,r}^{~2} \bigg\} ,
\end{split}
\label{H:exp}
\end{equation}

\noindent and

\begin{equation}
\begin{split}
  \mathcal{H} = & ~3 f^{\prime\prime} \left[ \left( g^{11}_{~,r} +
	g^{11} \left( \ln{\sqrt{-g}} \right)_{,r} \right) R_{,r}
	+ g^{11} R_{,rr} \right] \\
       & + 3 f^{\prime\prime\prime} g^{11} R_{,r}^{~2}. 
\end{split}
\label{curlH:exp} 
\end{equation}

  Under the assumption of spherical symmetry the angular terms of the
metric are not perturbed and so: 

\begin{equation}
	\sqrt{-g} = r^2 \sin{\theta} \left\{ 1 +
     \left[ g_{00}^{(2)} - g_{11}^{(2)} \right] +
     \mathcal{O}(4) \right\} ^{1/2},
\label{sqrt:g}
\end{equation}

\noindent then, by using the fact that
\( \ln\left(\sqrt{-g}\right)_{,r} = \left(\sqrt{-g}\right)_{,r} / \sqrt{-g} \),
it follows that

\begin{equation}
  \ln\left(\sqrt{-g}\right)_{,r} =  \frac{2}{r} + \frac{1}{2}
    \left[g_{00,r}^{(2)} - g_{11,r}^{(2)}\right] + \mathcal{O}(4) \, .
\label{log:g}
\end{equation}

  Since Ricci's scalar depends on the metric components and their
derivatives up to the second order with respect to the coordinates, it
follows it can only have a non-null second and higher
perturbation orders, i.e.

\begin{equation}
	R = R^{(2)} + R^{(4)} + \mathcal{O}(6) \, .
\label{ricci:exp}
\end{equation}

\noindent The fact that \( R^{(0)}=0 \) is consistent with the flatness
of spacetime assumption at the lowest zeroth order of perturbation.
The expression for the second order component of Ricci's scalar from
the metric components~\eqref{metric-perturbed} is given by

\begin{equation}
	R^{(2)} = - \frac{2}{r} \left[ g_{11,r}^{(2)} +
	\frac{g_{11}^{(2)}}{r} \right] - g_{00,rr}^{(2)} - \frac{2}{r}
	g_{00,r}^{(2)} \, .
\label{ricci:second}
\end{equation}

\noindent The global minus sign that appears on the right-hand side
of equation~\eqref{ricci:second} for Ricci's scalar \(R^{(2)}\)
at second perturbation order differs from that reported by
\citet{capozziello-newton,stabile09}. As mentioned above, and discussed in
appendix~\ref{appendix1}, this fact occurs due to the choice of signs in
the definition of Riemann's tensor.  The particular choice used throughout
the article is the one given by equation~\eqref{riemanndef} and so, our
solutions lie in a different branch as the one reported by those authors.

\section{Lowest order solution}
\label{lowest-order}

  Let us now calculate the order of the trace
equation~\eqref{trace:einstein} using relations~\eqref{eq3b}
and~\eqref{ricci:exp}. On the one hand,  the lowest order of the first two
terms on the left-hand side of the trace equation is \( \mathcal{O}(2b)
\). On the other hand, direct inspection of the right-hand side of
equation~\eqref{curlH:exp} results in the fact that the lowest order of \(
\mathcal{H} \) is \( \mathcal{O}(2b-2) \).  Indeed, the last term of the
right-hand side of this equation is \( \propto f^{\prime\prime\prime}
g^{11} R_{,r}^{~2} \) and so, to lowest order of perturbation of
relations~\eqref{metric:exp-up} and~\eqref{ricci:exp}, this means that \(
\mathcal{H} \) contains terms of the form \(  R^{(2)b-3} {}R^{(2)2}_{,r}
\) and so, \( \mathcal{H} \) is of order \( \mathcal{O}(2b - 2)\). This
analysis indicates that to lowest order the trace equation to consider is

\begin{equation}
  \mathcal{H}^{(2b - 2)} = 3 \Delta f^{\prime (2b-2)} (R) = 0 .
\end{equation}

\noindent This result is consistent with relation~\eqref{null:deltaR}
to lowest order of approximation and is in perfect agreement with the
perturbative study performed by \citet{bernal11}. Note also that
this is the only independent equation at this order.

  Direct substitution of equations~\eqref{eq3b} and \eqref{ricci:exp}
into the last equation leads to

\begin{equation}
\begin{split}
  & \mathcal{H}^{(2b - 2)} = 3 b(b -1) {}R^{(2)b - 2} g^{11(0)} \left[ 
    \left(\ln{\sqrt{-g}}\right)_{,r}^{(0)} R^{(2)}_{,r} \right. \\
    & \left. + R^{(2)}_{,rr} \right] + 3 b(b-1)(b -2) R^{{(2)}b - 3} 
    g^{11(0)} R_{,r}^{(2)2} = 0.
\end{split}
\label{h:step1}
\end{equation}

\noindent Substitution of 
expressions~\eqref{metric:exp-up} and~\eqref{log:g}
in the previous equation leads to the following differential equation for
Ricci's scalar at order \( \mathcal{O}(2) \):

\begin{equation}
  R^{(2)} \left[\frac{2}{r} R^{(2)}_{,r} + R_{,rr}^{(2)}\right] +
  (b -2)R_{,r}^{{(2)}2} = 0 \, ,
\label{eq1:notes}
\end{equation}

\noindent which can be written in a more suitable form as

\begin{equation}
  \left[ \ln R^{(2)}_{,r} \right]_{,r} + (b-2) \left[ \ln R^{(2)}
    \right]_{,r} = - \frac{2}{r} \, .
\end{equation}

\noindent The solution of the previous equation is:

\begin{equation}
  R^{(2)}(r) = \left[ (b-1) \left(\frac{A}{r} + B\right) 
    \right]^{1/(b-1)},
\label{R2:exact}
\end{equation}

\noindent where \(A\) and \(B\) are constants of integration.

  Far away from the central mass, spacetime is flat and so
Ricci's scalar must vanish at large distances from the origin.
This means that the constant \(B=0\) and so

\begin{equation}
  R^{(2)}(r) = \left[ (b-1) \frac{A}{r} \right]^{1/(b-1)}.
\label{R2:exactb}
\end{equation}

\noindent As explained by \citet{bernal11}, the case \( b = 3/2 \) yields a
MOND-like weak field limit and so, substituting \( b = 3/2 \) in
relation~\eqref{R2:exactb} yields:

\begin{equation}
\label{R2}
	R^{(2)}(r)  = \frac{\hat{R}}{r^2},
\end{equation}

\noindent where \(\hat{R}:= A^2/4\). This is exactly the same result
as the one obtained by \citet{bernal11}. As these authors have shown,
this result yields a MONDian-like behaviour for the gravitational field
in the limit \( r \gg l_M \gg r_\text{g} \).  For this particular case,
the lowest order of approximation of the theory is \( \mathcal{O}(1) \),
which has a higher relevance as compared to the order \( \mathcal{O}(2)
\) of standard general relativity for which \( b = 1 \). Using very
general arguments, the authors also showed that the constant \(\hat{R}
\propto r_\text{g}/l_M \) and so, \( \hat{R} \) is proportional to the
square root of the mass of the central object. In order to calculate \(
\hat{R} \) from perturbation analysis we need to find the expressions
for the metric at order \( \mathcal{O}(2) \) of approximation.

\section{$f(\chi) = \chi^{3/2}$ metric components}
\label{metric-components}

  Let us now solve the field equations at the next order \(\mathcal{O}(2b)\)
of approximation. At this order we expect the metric components
\( g_{00}^{(2)} \), \( g_{11}^{(2)} \) and Ricci's scalar \( R^{(4)}\)
to play a relevant role in the description of the gravitational field.
In fact, the field equations~\eqref{einstein:eqs} at this order are given by

\begin{equation}
  b R^{(2) b -1} R_{\mu\nu}^{(2)} - \frac{1}{2}  R^{(2) b}
g_{\mu\nu}^{(0)} + \mathcal{H}^{(2b)}_{\mu\nu} = 0 ,
\label{eq7a:notes}
\end{equation}

\noindent where

\begin{equation}
  \mathcal{H}^{(2b)}_{\mu\nu} = - \left( \nabla_\mu \nabla_\nu
     - g_{\mu\nu} \Delta \right) f^{\prime (2b)}(R).
\end{equation}

\noindent The complete \( \mathcal{H}^{(2b)}_{\mu\nu} \) from 
equation~\eqref{H:exp} is written in appendix~\ref{appendix3}.

  Now, from equation~\eqref{null:deltaR} it follows that the
Laplace-Beltrami operator applied to \(f^\prime (R)\) must be zero at
all perturbation orders. In particular \( \Delta f^{\prime (2b)} = 0 \). With
this condition, the field equations~\eqref{eq7a:notes} simplify greatly and 
can be written as

\begin{equation}
\begin{split}
  & b R^{(2) b -1} R_{\mu\nu}^{(2)} - \frac{1}{2}  R^{(2) b}
    g_{\mu\nu}^{(0)}   \\
  - & b (b-1) \bigg\{ R^{(2)b-2} \left[ R^{(4)}_{\mu\nu} -
    \Gamma^{1(0)}_{\mu\nu} R^{(4)}_{,r} - \Gamma^{1(2)}_{\mu\nu}
    R^{(2)}_{,r} \right] \bigg.   \\
  \bigg. + & (b-2) R^{(2)b-3} R^{(4)} \left[ R^{(2)}_{\mu\nu} - \Gamma^{1(0)}_{\mu\nu}
    R^{(2)}_{,r} \right] \bigg\}  \\
  - & b (b-1) (b-2) \bigg[ 2 R^{(2)b-3} R^{(2)}_{,\mu} R^{(4)}_{,\nu} \bigg.  \\
  \bigg. + & (b-3) R^{(2)b-4} R^{(4)} R^{(2)}_{,\mu} R^{(2)}_{,\nu} \bigg] = 0 .
\end{split}
\label{Hreduced:2m}
\end{equation}

\noindent Direct substitution of the following Christoffel symbols

\begin{equation}
  \Gamma^{1(0)}_{00} = 0 \, , \qquad \Gamma^{1(2)}_{00} = - \frac{1}{2}
    g^{11(0)} g^{(2)}_{00,r} \, , 
\end{equation}

\noindent and relations~\eqref{metric-perturbed} and~\eqref{metric:exp-up}
in the \(00\) component of equation~\eqref{Hreduced:2m} leads to

\begin{equation}
  b R^{(2)b-1} R_{00}^{(2)} - \frac{1}{2} R^{(2)b}  +  \frac{1}{2}
    b (b-1) g^{(2)}_{00,r} R^{(2)b-2} R^{(2)}_{,r} = 0 .
\label{eq11a:notes}
\end{equation}

\noindent If we now substitute \( b = 3/2 \), expression~\eqref{R2} and the
value of Ricci's tensor at \( \mathcal{O}(2) \) of approximation:

\begin{equation}
  R^{(2)}_{00} = - \frac{r g^{(2)}_{00,rr} + 2 g^{(2)}_{00,r}} {2r},
\label{ricci:00}
\end{equation} 

\noindent into equation~\eqref{eq11a:notes}, we obtain the following
differential equation for \( g^{(2)}_{00} \):

\begin{equation}
  r^2 g^{(2)}_{00,rr} + 3 r g^{(2)}_{00,r} + \frac{2 \hat{R}}{3} = 0 ,
\label{gtt2}
\end{equation} 

\noindent and so

\begin{equation}
	g^{(2)}_{00}(r) = - \frac{\hat{R}}{3}  \ln \left( \frac{r}{r_\star} \right)
	+ \frac{ k_1 }{ r^2 } , 
\label{g00:notes}
\end{equation}

\noindent where \(k_1\) and \(r_\star\) are constants of integration.
By substitution of this result in equation~\eqref{ricci:second} and using
equation~\eqref{R2} we get the following differential equation
for \(g^{(2)}_{11}\): 

\begin{equation}
  r g_{11,r}^{(2)} + g_{11}^{(2)} + \frac{k_1}{r^2} 
    + \frac{\hat{R}}{3} = 0 ,
\label{grr2:diff}
\end{equation}

\noindent with solution:

\begin{equation}
  g^{(2)}_{11}(r) = \frac{k_1}{r^2} + \frac{k_2}{r} - \frac{\hat{R}}{3} ,
\label{grr2}
\end{equation}

\noindent where \(k_2\) is a constant of integration.

\section{Metric coefficients from astronomical observations}
\label{metric-coefficients}

  In this section we derive the constraints that the well established
astrophysical phenomenology of asymptotically flat galactic rotation
curves satisfying the Tully-Fisher relation, and the cumulative
gravitational lensing observations for elliptical and spiral galaxies and 
galaxy groups, imply for the metric coefficients for static, spherically
symmetric spacetimes for any metric theory of gravity where dark matter
is not required.

  To begin with, let us take the radial
component~\eqref{radial:geodesic} of the geodesic
equations~\eqref{geodesic} in the weakest limit of the theory.  In this
limit, the rotation curve for test particles bound to a circular
orbit about a mass \(M\) with circular velocity \( v(r) \) given by
equation~\eqref{centrifugal} is

\begin{equation}
   \frac{v^{2}(r)}{c^{2}r} = \frac{1}{2} g^{11} g_{00,r}.
\label{v:metric}
\end{equation}

\noindent Except for the inner regions of spiral galaxies, \(v(r)\)
can be well approximated by a constant which scales with the fourth
root of the total baryonic mass \(M_\text{b}\) of the spiral galaxy in
question, as described by the Tully-Fisher empirical relation \citep[see
e.g.][]{milgrom08,famaey11}

\begin{equation}
  v = (G M_\text{b} a_{0} )^{1/4}.
\label{tully-fisher}
\end{equation}

\noindent In fact, it is from numerous observations of galactic rotation
curves and total baryonic mass estimates, that the constant \(a_{0}\)
has been calibrated \citep[see e.g.][and references therein]{famaey11}.

  We now substitute equations~\eqref{metric-perturbed}
and~\eqref{metric:exp-up} to order \(\mathcal{O}(2)\) of approximation and
relation~\eqref{tully-fisher} in equation~\eqref{v:metric} to
obtain the following differential equation for \(g_{00}^{(2)}\):

\begin{equation}
  - g^{(2)}_{00,r} = \frac{2}{r} \left( \frac{v}{c} \right)^2 
     = \frac{2 (G M_\text{b} a_0)^{1/2}}{c^{2} r} ,
\end{equation}

\noindent having as solution

\begin{equation}
\begin{split}
  & - g^{(2)}_{00}(r) = 2 \left( \frac{v}{c} \right)^2
    \ln \left( \frac{r}{r_\star} \right)   \\
  & = \frac{ 2 ( G M_\text{b} a_0 )^{1/2} }{ c^{2} } \ln
    \left( \frac{r}{r_\star} \right) = \frac{2 r_{\text g}}{l_M}
    \ln \left( \frac{r}{r_\star} \right) ,
\end{split}
\label{g00:empirical}
\end{equation}

\noindent where \(r_\star\) is a scale radius which, from the point of
view only of the flat rotation curves of galaxies and the Tully-Fisher
relation, remains arbitrary. We therefore see that a necessary and
sufficient condition in any metric relativistic theory of gravity, where
all observational constraints of galactic rotation curves are satisfied
without invoking dark matter, is that \(g^{(2)}_{00}\) must
satisfy the previous empirically derived relation.

  Comparing the theoretical metric coefficient
\(g^{(2)}_{00}\) given by~\eqref{g00:notes} (obtained from perturbation
theory for \(f(\chi)=\chi^{3/2}\))
and the empirical one~\eqref{g00:empirical} (obtained from the phenomenology of
flat rotation curves and the Tully-Fisher relation), give the following
values for the integration constants needed in equation~\eqref{g00:notes}:

\begin{equation}
  k_1 = 0, \qquad \hat{R} =  6 r_\text{g} / l_M.
\label{s04}
\end{equation}

  In this case, the gravitational potential \( \phi \)
from equation~\eqref{weakest-metric} takes the form:

\begin{equation}
  \phi = - v^2 \ln \left( \frac{r}{r_\star} \right) =
     - \left( G M_\text{b} a_0 \right)^{1/2} \ln \left(
     \frac{r}{r_\star} \right),
\label{s03}
\end{equation}

\noindent which yields a radial MONDian acceleration:

\begin{equation}
  | a | =  | \nabla \phi | =  \frac{ \left( G M_\text{b} a_0 \right)^{1/2} }{r},
\label{mondian-acceleration}
\end{equation}

\noindent  Thus, in the \( v/c \ll 1 \) limit, the \( f(\chi) =
\chi^{3/2} \) presented is seen to agree with the observed phenomenology
of the observed galactic rotation curves in the absence of dark matter,
as already shown by \citet{bernal11}.

  The \(g_{11}\) metric coefficient will be obtained from gravitational
lensing phenomenology. We begin from the general deviation angle equation
written from the point of view of an observer at infinity, the ``astronomer''
detecting the gravitational lens in question \citep[see
e.g.][]{weinberg,schneider,keeton05}:

\begin{equation}
  \beta = 2 \int_{r_\text{i}}^{\infty}  \frac{[-g_{00}(r) g_{11}(r)]^{1/2}
    \text{d}r}{r\left[(r/r_\text{i})^{2} g_{00}(r_\text{i})  - g_{00}(r)
    \right]^{1/2} } - \pi,
\label{beta}
\end{equation}

\noindent where \(r_\text{i}\) is the closest approach to the central mass M,
and it is related to the impact parameter \( b \) through the relation 
\( r_\text{i}^2 = b^2 \, g_{00}(r_\text{i}) \). 

  Over the last few years it has become clear that the complete
phenomenology of gravitational lensing, at the level of extensive massive
elliptical galaxies \citep[see e.g.][]{gavazzi07,koopmans06,barnabe11},
galaxy groups \citep[see e.g.][]{more11}, clusters of galaxies \citep[see
e.g.][]{newman09,limousin07} and more recently spiral galaxies \citep[see
e.g.][]{dutton11,suyu12} can be accurately modelled using total matter
distributions having isothermal profiles, when treating the problem from
the point of view of Einstein's general relativity. All these observations
show that the dark matter halos needed to explain gravitational lensing
under Einstein's general relativity obey the same Tully-Fisher scaling
with total baryonic mass as the ones needed to explain the observed
rotation curves of spiral galaxies. This means that for a given total
baryonic mass, spiral and  elliptical galaxies and groups of galaxies
require dark matter halos having the same physical properties to explain
the observations; from kinematics of rotation curves in the former case
to gravitational lensing in the latter one \citep{dutton11,suyu12}.
Under Einstein's general relativity the majority of these isothermal
matter distribution, particularly at large radii, must be composed of
a hypothetical dark matter.

  For a static spherically symmetric total matter distribution \(
M_\text{T} \), since assuming the validity of Einstein's general
relativity Schwarzschild's metric holds, and therefore \(g_{00\text{S}}
= - 1/g_{11\text{S}}\), we get:

\begin{equation}
  g_{00\text{S}} = 1 - \frac{2 r_\text{g}}{r}
    = 1- \frac{2 G M_\text{T}(r)}{c^{2} r}
    = 1- 2 \left( \frac{v}{c} \right)^2.
\label{g00S}
\end{equation}

\noindent The subscript S identifies the coefficients of the Schwarzschild
metric, and \(M_\text{T}(r)=v^{2}r/G\) refers to the hypothetical
isothermal total matter distribution \citep[cf.][]{binney-tremaine} needed to
explain the observed lensing, when assuming general relativity. From this
it follows that the dark matter hypothesis provides a self-consistent
interpretation of observed phenomenology: the same 
dark matter halos, which are required to explain the observed rotation
curves, have been solved for by analysing extensive lensing observations.

  From equation~\eqref{g00S} it follows that for isothermal total
matter halos under Einstein's general relativity, the metric coefficient
\(g_{00\text{S}}\) does not depend on the radial coordinate. We can see
this by using the empirical Tully-Fisher relation~\eqref{tully-fisher}
between the velocity and the total baryonic mass in the last identity
above. Thus, the coefficient~\eqref{g00S} can then be taken outside
of the integral~\eqref{beta} of the deviation angle, where for the
Schwarzschild metric and isothermal total matter halos we now obtain

\begin{equation}
  \beta = \frac{2}{[1-2(v/c)^{2}]^{1/2}} \int_{r_\text{i}}^{\infty} \frac
    {\text{d}r}{r\left[(r/r_\text{i})^{2} - 1 \right]^{1/2}} - \pi .
\label{sergito03}
\end{equation}

\noindent The above radial integral yields \(\pi/2\) and we obtain the
observed bending angle as

\begin{equation}
  \beta=\frac{\pi}{ [1-2(v/c)^{2}]^{1/2} } - \pi = \frac{\pi}
    { [1-2(G M_\text{b} a_{0})^{1/2}/c^{2}]^{1/2} } - \pi .
\label{sergito01}
\end{equation}

  We see that the well established empirical result of lensing
observations yielding isothermal total dark matter halos under the
standard theory is strictly the observation of constant bending angles
which do not depend on the impact parameter, scaling with the observed
baryonic total masses as indicated above.

  Now, since \( (v/c)^2 \) is of order \( \mathcal{O}(2) \) we can write
equation~\eqref{sergito01} as

\begin{equation}
  \beta= \pi \left( \frac{v}{c} \right)^{2}  = \pi \frac{(G M_\text{b}
    a_{0})^{1/2}}{c^{2}} = \pi \frac{r_\text{g}}{l_M} .
\label{sergito02}
\end{equation}

\noindent The above equation summarises all empirical results of gravitational
lensing at galactic and galaxy group scales: the bending angle does
not depend on the impact parameter and scales with the square root of
the total baryonic mass. This last equation gives a clear illustration
of the link between the dynamics and the spacetime curvature effects
induced by the presence of an observed baryonic mass.

  We can now use the result of equation~\eqref{sergito02} to constrain the
metric coefficient \(g_{11}\) for any metric theory of gravity, seeking
an accurate description of the observed gravitational lensing phenomena
without the introduction of any hypothetical dark matter. To do this,
let us return to the general lensing equation~\eqref{beta}, and ask that
the results obtained under the Schwarzschild metric with isothermal total
matter halos match those under any metric theory of gravity,
at all impact parameters and for any total baryonic masses:

\begin{equation}
\begin{split}
    \frac{1}{2} \left( \beta + \pi \right) & = \left[ 1 + \left(
    \frac{v}{c} \right)^2 \right] \int_{r_\text{i}}^{\infty}
      \frac{\text{d}r}{r \left[ (r/r_\text{i})^{2} - 1 \right]^{1/2}}  \\
    & = \int_{r_\text{i}}^{\infty} \frac{ \left[ - g_{00}(r) g_{11}(r)
      \right]^{1/2} \text{d}r}{r \left[ (r/r_\text{i})^{2} g_{00}(r_\text{i})
      - g_{00}(r) \right]^{1/2} },
\end{split}
\label{sergito04}
\end{equation}

\noindent at \( O(2) \) of approximation from equations~\eqref{beta}
and~\eqref{sergito03}.  Let us rearrange integral~\eqref{sergito04} in such
a way that:

\begin{equation}
\begin{split}
    & \int_{r_\text{i}}^{\infty} \left\{ \left[ 1 + \left( \frac{v}{c}
       \right)^2 \right] \frac{1}{r \left[(r/r_\text{i})^{2}
       - 1 \right]^{1/2}} \right. \\
    & \left. - \frac{\left[ - g_{00}(r) g_{11}(r) \right]^{1/2}}
      {r \left[ (r/r_\text{i})^{2} g_{00}(r_\text{i})- g_{00}(r) \right]^{1/2} }
      \right\} \text{d}r = 0 .
\end{split}
\end{equation}

\noindent Since the result must hold for all impact parameters, the
integrand of the above equation must be equal to zero and so,

\begin{equation}
  \left[ 1+ \left( \frac{v}{c} \right)^2 \right]^{2}
    \frac{1}{(r/r_\text{i})^{2}-1} = \frac{- g_{00}(r)
    g_{11}(r)}{(r/r_\text{i})^{2}g_{00}(r_\text{i}) - g_{00}(r)  },
\end{equation}

\noindent  Approximating the previous relation to order \( \mathcal{O}(2) \) 
it follows that that the metric coefficient \( g_{11} \) is given by:

\begin{equation}
  g_{11} (r)= - \left[ 1 + 2 \left( \frac{v}{c} \right)^2
     \right] \frac{(r/r_\text{i})^{2} \left[ g_{00}(r_\text{i})
     / g_{00}(r) \right] - 1}{(r/r_\text{i})^{2} - 1} .
\label{sergito05}
\end{equation}

  From a mathematical point of view, since the contribution to the integral
in the lensing equation~\eqref{beta} is fully dominated by the region
\(r \approx r_\text{i}\), and given the very mild radial dependence of the
empirical \( g_{00} \) term, we can take \( g_{00}(r_\text{i}) \approx g_{00}(r) \)
in the above equation to yield:

\begin{equation}
\begin{split}
  g_{11}(r) & = - 1 - 2 \left( \frac{v}{c} \right)^{2}  \\
    & = - 1 - \frac{2 (G M_\text{b} a_0)^{1/2}}{c^{2}}
      = - 1 - 2 \frac{r_\text{g}}{l_{M}} .
\end{split}
\label{g11:empirical}
\end{equation}

  Thus, any metric theory of gravity where \(g_{11}\) matches the
above expression in the regime where gravitational lenses are observed
will accurately reproduce all the observed lensing phenomenology,
with \( M_\text{b} \) the total baryonic mass of the object in question
(galaxies or group of galaxies), and no hypothetical dark matter assumed
to exist. Equations~\eqref{g00:empirical} and~\eqref{g11:empirical}
give empirical mathematical relations for the metric coefficients at
perturbation order \( \mathcal{O}(2) \) which reproduce all observed
rotation velocity and gravitational lensing phenomenology, without the
inclusion of any dark matter component.

  Notice that the mass dependence of the second term on the right-hand side 
in expression~\eqref{g11:empirical} for the metric coefficient \(g_{11}\) 
is the same as the factor in 
expression~\eqref{g00:empirical} for \(g_{00}\).  This last was
obtained for a rigorously flat rotation curve in
accordance with the Tully-Fisher relation. This shows that
the ratio \(r_\text{g}/l_{M}\) of the two important characteristic lengths
of the extended metric theory of gravity proposed by \citet{bernal11} is the
determinant dimensionless measure of deviations from flat spacetime
at galactic scales, exactly as expected from the dimensional analysis in
\citet{hernandez12}

  The metric coefficient \( g_{11} \) in equation~\eqref{g11:empirical} can
be directly compared to the results for the \( f(\chi) = \chi^{3/2} \)
metric theory of \citet{bernal11} obtained in equation~\eqref{grr2} with
the inclusion of the results of equation~\eqref{s04}.  This means that the
choice of the integration constant

\begin{equation}
  k_2 = 0,
\label{sergio06}
\end{equation}

\noindent makes these expressions for the metric component \( g_{11} \) 
identical.

  Use of the mathematical approximation \( A^{x} \approx 1 + x \ln A \) to
write the following expressions for the full empirical metric coefficients
gives:

\begin{gather}
g_{00} \approx 1 + \left( 2 r_\text{g} / l_M \right) \ln\left( r_\star / r
     \right) \approx \left( r_\star / r \right)^{ 2 r_\text{g} / l_M }, 
      \label{hola1} \\
g_{11} \approx  - 1 - \left( 2 r_\text{g} / l_M \right)
  \approx e^{ 2 r_\text{g} / l_M }.
      \label{hola2} 
\end{gather}

  We note that all the approximations used in this section introduce
an error several orders of magnitude smaller than the intrinsic observational
uncertainties in the empirical relations used. Therefore, all of the
expressions given can be considered as strictly equivalent in regards to
the accurate modelling of astrophysical rotation curves and gravitational
lensing data.

\section{Discussion}
\label{discussion}

  Through the use of the weak field limit of the metric \( f(\chi)
= \chi^{3/2} \) theory of gravity constructed by \citet{bernal11},
we have shown that it is possible to explain both the dynamics of
massive particles and the deflection of light by observed astronomical
systems such as elliptical galaxies, spiral galaxies and groups
of galaxies.  Recently, the same metric theory of gravity was shown
to be coherent also with the expansion dynamics of the observed
universe~\citep{carranza12,mendoza12}.  This is an expected result
from a theory of gravity constructed through astronomical observations:
it must be coherent at all scales.   The regime of Einstein's general
relativity is by no means violated, since the applications developed in
this article (\( r \gg l_M \)) lie far away from the mass and length
scales associated to the ones of Einstein's general relativity (\(
r \ll l_M \)) \citep[see e.g.][]{mendoza12}.

  The results of this article were constructed using a static
spherically symmetric metric with the time and radial components
perturbed up to order \(\mathcal{O}(2)\) of approximation. This work
generalises the one of \citet{bernal11} in which the radial metric
component was assumed up to order \(\mathcal{O}(0)\) only and so,
information on the choice of signature of the Riemann tensor was lost
(see appendix~\ref{appendix1}). Such information is very important while
working with fourth order metric theories of gravity.

  We mention again the tremendous importance of a correct choice for the
signature of the Riemann tensor as described in appendix~\ref{appendix1}.
The choice~\eqref{riemanndef}, and only that choice, used in this
article yields results in agreement with astronomical observations.
In other words, astronomical observations fix the correct (and unique)
choice of signature for Riemann's tensor.  This is an important result,
since otherwise solutions from the other branch appear which are not in
accordance with astronomical observations.

  Table~\ref{table01} summarises our main results.  It is important to
note that the empirical values of the metric components \( g_{00}^{(2)}
\) and \( g_{11}^{(2)} \) do not depend on any gravitational theory and
as such, they represent functions that any successful theory of gravity
(such as the one used in this article) needs to match.  Notice that
observationally, independent empirical constraints fixed the \( 2 /
r_\text{g} / l_M \) factors in \( g_{00} \) and \( g_{11} \) to be equal;
it is encouraging that the formal mathematical perturbation treatment
of the theory proposed also yields identical \( \hat{R} / 3 \) factors
in the expressions for \( g_{00} \) and \( g_{11} \).  If this were
not the case, even given the compatible functional forms of empirical
and theoretical metric coefficients, the \( f(\chi) = \chi^{3/2} \)
proposal would have been rejected.


\begin{table}
\begin{center}
\begin{tabular}{@{}|l|c|c|}
\hline \hline & &\\
Metric & \(g_{00}^{(2)} \) & \( g_{11}^{(2)} \)   \\
coefficient & & \\
             &  & 					\\
\hline       &  &  					\\
             & \( - \frac{2 r_\text{g}}{l_M} \ln \left( \frac{r}{r_\star}
	     \right) \) &
                     \( - \frac{2r_\text{g}}{l_M} \)  		\\
Observations &   & 					\\
             &   (Tully-Fisher)      &  (lensing)       \\
             &   & 					\\
\hline       &   & 					\\
             & \(-\frac{\hat{R}}{3}\ln \left( \frac{r}{r_\star} \right) +
\frac{k_1}{r^2}\)
& \( \frac{k_1}{r^2} + \frac{k_2}{r} - \frac{\hat{R}}{3} \) \\
Theory  &                                     &
\\
\( f(\chi) = \chi^{3/2} \)      &  \( \hat{R} =  6 r_\text{g} / l_M \) \   \( k_1 = 0 \)     &
\( \hat{R} =  6 r_\text{g} / l_M  \ k_2 = 0  \) \\
             &                                                          &
\\
\hline \hline
\end{tabular}
\end{center}
\caption[Observations vs theory results]{The table shows the results
obtained for the metric components \( g_{00}^{(2)} \) and \( g_{11}^{(2)}\)
for a static spherical symmetric spacetime in scales of galaxies 
and galaxy groups obtained empirically from astronomical observations of these
systems and the ones predicted by the metric \( f(\chi) = \chi^{3/2} \)
theory of gravity of \citet{bernal11}. A good metric theory of gravity 
must be such that it converges to the inferred values presented in the
table. The theory \( f(\chi) = \chi^{3/2} \) is in perfect agreement with
the observed metric components.  The dimensionless ratio
formed by the quotient of the gravitational radius \( r_\text{g} \) to the
mass-length scale \( l_M \) (see equation~\eqref{defs}) is the determinant
dimensionless quantity of the problem.  Since the metric components
determine the ``gravitational potential'' of the system, the length \(
r_\star \) is undetermined.  However, since the natural length scale of the
system is \( l_M \) one can always assume \( r_\star = l_M \), which also
ensures no sign change in the potential in equation~\eqref{s03} over the
domain of applicability \( r > l_M \).
}
\label{table01}
\end{table}


  An important fact arises from the usage of the \( f(\chi) \) metric
theory of gravity and not the \( f(R) \) formalism.  Although closely
related to each other for a power-law function~\eqref{eq3b} and a mass
point source, the correct dimensional approach \( f(\chi) \) introduces
mass and length scales that, as shown by \citet{bernal11}, need to
be incorporated into the gravitational field action.  Although the
field equations in vacuum for both \( f(\chi) \) and \( f(R) \) under a
power-law representation yield the same field equations (since the mass \(
M \) generating the gravitational field is a constant), \( f(R) \) gravity
is not capable of reproducing the crucial lensing observations as it lacks
a crucial constraint equation~\eqref{null:deltaR}.  The gravitational
theory \( f(\chi)=\chi^{3/2} \) is able to do so since under this
approach the correct limit where MONDian-like effects are expected
yield the constraint equation~\eqref{null:delta} or~\eqref{null:deltaR}.
Notice however that both \( f(R) \) and \( f(\chi) \) with the appropriate
choice of Riemann's tensor~\eqref{riemanndef} are able to reproduce the
flat rotation curves of galaxies and the Tully-Fisher relation.

  In an effort to generalise and look for a fundamental basis to an \(
f(\chi) \) theory of gravity, \citet{carranza12} and \citet{mendoza12}
have shown that these metric theories are equivalent to the the \(F(R,T)\)
construction of \citet{harko11}.  These authors have also shown that the
particular theory \(f(\chi)=\chi^{3/2}\) is in excellent agreement with
cosmological observations of SNIa.

  An \( f(\chi) \) theory of gravity satisfying the limits of
equation~\eqref{f-chi-real} implies that gravity is no longer
scale-invariant.  In fact, precise gravity tests have been performed
only at strong regimes of Einstein's gravity, where \( \chi \gg 1 \),
and so the involved accelerations of test particles are such that \(
a \gg a_0 \) \citep[see e.g.][]{will06}.  In exactly the opposite
regime, where \( \chi \ll 1 \), where the involved accelerations of
test particles are such that \( a \lesssim a_0 \), gravity differs
from Einstein's general relativity.  The traditional approach of
assuming Einstein's general relativity to be valid at all scales
means that unknown dark entities are needed to explain various
astrophysical observations.  This article heavily reinforces many others
\citep{bernal11,bernal11b,carranza12,mendoza11,mendoza12,hernandez10,hernandez12a,
hernandez12b} that show how astrophysical and cosmological observations
can be accounted for without assuming the existence of dark entities
and extending gravity so as to be non scale-invariant.

\section{Acknowledgements}
\label{acknowledgements}

  The authors acknowledge the input of an anonymous referee, helpful
in reaching a clearer final version of the article. This work was
supported by three DGAPA-UNAM grants (PAPIIT IN116210-3, IN111513-3 and
IN103011-3). The authors TB, XH, JCH, SM \& LAT acknowledge economic
support from CONACyT: 207529, 25006, 51009, 26344 and 221045. The
authors thank the kind assistance provided by Cosimo Stornaiolo for
finding the analytic solution~\eqref{R2:exact} of the differential
equation~\eqref{eq1:notes}.

\bibliographystyle{mn2e}
\bibliography{lensing-fchi}

\appendix

\section{Comments about the sign convention in Riemann's tensor}
\label{appendix1}

  In the study of the gravitational field equations, 
the link between the curvature of spacetime and the matter content is a key
fact. All
the information regarding the curvature of spacetime is contained
in the Riemann curvature tensor \( R^{\alpha}_{\ \beta\eta\theta} \),
which is a function of the first and second derivatives of the metric.
From a purely mathematical point of view, the Riemann tensor can be
obtained from the Com-mutator of covariant derivatives \citep{carroll04}:

\begin{equation}
  [\nabla_{\mu},\nabla_{\nu}] V^\rho = R^{\rho}_{\ \sigma\mu\nu} V^{\sigma},
\label{eqtempo2}
\end{equation}

\noindent for any vector field \( V^\alpha \).  From a geometrodynamical
point of view, the curvature tensor is constructed through the change
\( \Delta A_\mu \) in a vector \( A_\mu \) after being  displaced
about any infinitesimal closed contour \citep{daufields}: \( \Delta
A_\mu = \oint{ \Gamma^\lambda_{\mu\nu} A_\lambda \mathrm{d} x^\nu } \).
By the use of Stokes' theorem it then follows that for a sufficiently small
closed contour:

\begin{equation}
  \Delta A_\mu \approx \frac{ 1 }{ 2 } R^\lambda_{\ \mu\nu\theta} A_\lambda \Delta
  f^{\nu\theta},
\label{eqtempo}
\end{equation}

\noindent where \( \Delta f^{\nu\theta} \) represents the infinitesimal
area enclosed by the contour of the line integral.  In this respect, it
follows that the Riemann tensor measures the curvature of spacetime
\citep[cf.][]{daufields}.

  In equations~\eqref{eqtempo2} and~\eqref{eqtempo}, the Riemann tensor has
been defined as:

\begin{equation}
  R^{\beta}_{\ \mu\nu\alpha} := \Gamma^{\beta}_{\ \mu\alpha,\nu} 
           - \Gamma^{\beta}_{\ \mu\nu,\alpha} 
           + \Gamma^{\beta}_{\ \lambda\nu}\Gamma^\lambda_{\ \mu\alpha}
           - \Gamma^{\beta}_{\ \lambda\alpha}\Gamma^\lambda_{\ \mu\nu}.
\label{riemanndef}
\end{equation}

\noindent If Riemann's tensor is defined by equation~\eqref{riemanndef},
then Ricci's tensor is \( R_{\nu\alpha} := g^{\beta\mu}
R_{\beta\mu\nu\alpha} \) and Ricci's scalar is \(
R^\alpha_{\phantom{\alpha}\alpha} \).  Since these are the most used
definitions in relativity theory nowadays, we will refer to these quantities
as ``\emph{standard}''.

  However, there is another way in which Riemann's tensor (and Ricci's
tensor) can be defined, usually adopted by mathematicians and by Computer
Algebra Systems (CAS) such as Maxima (http://maxima.sourceforge.net).
In these cases, the syntaxis is such that \cite[see e.g.][]{toth05}

\begin{equation}
\begin{split}
  R[\mu,\nu,\alpha,\beta] &:= R^{\beta}_{\,\,\mu\nu\alpha}  \\
       & = \Gamma^{\beta}_{\,\,\mu\nu,\alpha} 
           - \Gamma^{\beta}_{\,\,\mu\alpha,\nu}
           + \Gamma^{\beta}_{\,\,\lambda\alpha}\Gamma^\lambda_{\,\,\mu\nu}
           - \Gamma^{\beta}_{\,\,\lambda\nu}\Gamma^\lambda_{\,\,\mu\alpha}.
\end{split} 
\label{riemannmax}
\end{equation}

\noindent  If Riemann's tensor is defined by equation~\eqref{riemannmax},
then Ricci's tensor is \( R_{\nu\alpha} := g^{\beta\mu}
R_{\beta\mu\nu\alpha} \) and Ricci's scalar is \( R^\alpha_{\ \alpha} \).
Although this choice of signs for the Riemann and Ricci tensors is not
very much in use these days, some well-known textbooks use them \citep[see
e.g. the Table of Sign Conventions at the beginning of reference][]{MTW}.

\noindent The CAS Maxima uses the definition~\eqref{riemannmax} and is such
that:

\begin{equation}
  \mathbf{R}_{\mathrm{maxima}} = - \mathbf{R}_{\mathrm{standard}} ,
\end{equation}

\noindent in free-index notation.

  As discussed in the Table of Sign Conventions of \citet{MTW},
general relativity can use any of the above definitions (and a few
more) simply because of the linearity with which Ricci's scalar and
Ricci's tensor appear in Einstein's field equations.  This is however not the
case in metric \( f(R) \) theories of gravity, since for example in
those theories, the trace of the field equations is given by \citep[see
e.g.][]{capozziello-book}:

\begin{equation}
  f^\prime(R)R - 2 f(R) + 3 \Delta f^\prime(R) =
    \frac{8 \pi G}{c^4} T.
\end{equation}

\noindent To highlight the point, let us substitute the power-law 
function~\eqref{eq3b} in the previous equation to obtain:

\begin{equation}
  (b-2) R^{b} + 3b \Delta R^{b-1} = \frac{8 \pi G}{c^4} T.
\end{equation}

\noindent This equation reflects a crucial fact about the choice of
sign in Riemann's tensor.  Due to the presence of the derivative term
\(f^\prime(R) = b R^{b-1}\), depending on the sign convention of the
definition of the Ricci scalar, there appears a sign factor \( (\pm)^{b-1}
\) which is not global to all the terms in the equation. This establishes
a \emph{bifurcation} in this class of solutions of the theory. Indeed,
for a situation where \( f(R) = R^a + R^b \) or any more complicated
function of \(R\), there is not (\emph{a priori}) any indication of
which convention in the definition of Riemann's tensor should be used
to describe a particular physical phenomena.  In this article we show
that, under the theory being presented, the convention can be settled.
The results presented in this article were obtained with the standard
definition of Riemann's tensor in equation~\eqref{riemanndef}.  That
choice (and only that one) can account for both observed dynamics of
massive particles  in spiral galaxies through the Tully-Fisher relation,
and for the deflection of light observed in gravitational lenses.
An important aspect to point out is that the case \( f(R) = R \)
of Einstein's general relativity is free from the above ambiguity.
This is so because it is possible to redefine the signature for the
energy-momentum tensor to recover the same field equations \citep[see
e.g.][]{MTW,hobson06}.

  We see from this result that previous works by
\citet{capozziello-newton,stabile09} have selected the convention used
by the CAS Maxima in order to compute their results.  In that respect,
their results lie in another branch of the solutions of the
field equations.  If we would have taken for example, the definition
of Riemann's tensor by Maxima, then the metric coefficients would have
been: \( g_{00}^{(2)} = 2 \hat{R} \ln^2( r ) / 9 + A \, \ln(r) + B \)
and \( g_{11}^{(2)} = -2 \hat{R} \ln( r ) / 9 + D / r + ( \hat{R} -
A )/2 \) (where \( A \), \( B \) and \( D \) are constants). These are
very different from the ones obtained in equations~\eqref{g00:notes}
and~\eqref{grr2} and would have never reproduced the astrophysical
observations treated in this article.  It is only through the correct
choice of signs in the definition of Riemann's tensor, such as the ones
used in the present article and represented in equation~\eqref{riemanndef},
that the good agreement with the Tully-Fisher relation and lensing
observations can be correctly obtained.

\section{Comments about the maxima code}
\label{appendix2}

  In this section we give a brief introduction to the code we wrote in 
the Computer Algebra System (CAS) Maxima (http://maxima.sourceforge.net)
to obtain the field equations. Specifically, we work with the module
\verb+ctensor+ \citep[cf.][]{toth05}. The syntax of such module is that, when
invoked, it runs an input interface to design the form of the covariant
metric.

  The Maxima code MEXICAS (Metric EXtended-gravity Incorporated through
a Computer Algebraic System) is Copyright of T. Bernal, S. Mendoza
and L.A. Torres, licensed under a GNU Public General License (GPL),
version 3 (see http://www.gnu.org/licenses) can be obtained from:
http://www.mendozza.org/sergio/mexicas (see the section about copyright
and usage in that webpage).

  For the implementation of the code, we consider a perturbative approach 
in the parameter \(\epsilon := 1/c\), such that the covariant components of the
metric are given by

\begin{equation}
\begin{split}
  g_{00} & = 1 + \epsilon^2 g_{00}^{(2)} + \mathcal{O}(4),  \\
  g_{11} & = -1 + \epsilon^2 g_{11}^{(2)} + \mathcal{O}(4), 
\end{split}
\end{equation}

\noindent where the angular components are given by the
standard expressions for spherical coordinates as shown in
equation~\eqref{metric-perturbed}.  With these equations, it is simple
to construct the contravariant components of the metric:

\begin{equation}
\begin{split}
  g^{00} & = 1 - \epsilon^2 g_{00}^{(2)} + \mathcal{O}(4),  \\
  g^{11} & = -1 - \epsilon^2 g_{11}^{(2)} + \mathcal{O}(4).
\end{split}
\end{equation}

  With these considerations, the metric is recorded in the \verb+ctensor+
module. From this fact, it is simple to invoke all the quantities required
to construct the field equations, either in general relativity or for
any extended metric theory of gravity. For example, in a descriptive
way concerning the syntaxis of maxima it follows that:

\begin{equation}
  \verb+christof(mcs)+ \longrightarrow
    \Gamma^{\lambda}_{\,\,\,\mu\nu} ,
\end{equation}

\noindent and with similar syntaxis for the Riemann tensor,
the Ricci tensor and the Ricci scalar.

  Due to the fact that the metric has an order parameter
\( \epsilon \), all the tensorial quantities involved in
the construction of the field equations will gain this
dependence. In the formalism of the code, it is a crucial
fact to extract the perturbation order of every metric quantity 
to construct the field equations at the desired perturbation order.
For example, for a generic quantity \(q\) calculated from the
manipulation of the metric, if we consider that \(q^{(n)}\)
represents such quantity at order \(n\), we have:

\begin{equation}
  q^{(0)} = \lim_{\epsilon \to 0} q,
\end{equation}

\noindent which reproduces the flat spacetime limit. For the second 
order we have

\begin{equation}
  q^{(2)} = \lim_{\epsilon \to 0} \frac{ q - q^{(0)} -
     \cancelto{0}{\epsilon q^{(1)}} }{ \epsilon^2},
\label{sergito07}
\end{equation}

\noindent and consequently the fourth order is obtained by

\begin{equation}
  q^{(4)} = \lim_{\epsilon \to 0} \frac{ q  - q^{(0)}
    -\cancelto{0}{\epsilon q^{(1)}} - \epsilon^2 q^{(2)} -
    \cancelto{0}{\epsilon^3 q^{(3)}} } {\epsilon^4}.
\label{sergito08}
\end{equation}

\noindent Similarly, higher perturbation orders can be obtained by the
obvious generalisation of the previous relation.

  In equations~\eqref{sergito07} and~\eqref{sergito08}, it is implied
that the first order quantities vanish, as is also the case for the
Christoffel symbols.  This computational procedure gives as an output
a key result used in the article corresponding to Ricci's
scalar at second perturbation order, given by equation~\eqref{ricci:second}.

\section{Extended field equations using Maxima}
\label{appendix3}

  By using the Computer Algebra System (CAS) Maxima and the MEXICAS code
(see appendix~\ref{appendix2}), we obtained the
field equations up to the second order.

  The trace~\eqref{trace:einstein} of the field equations~\eqref{eq7a:notes}
to the order \(\mathcal{O}(2b)\) of approximation
can be simplified  with the aid of the solutions found at the lowest order of
approximation in Section~\ref{lowest-order} to obtain

\begin{equation}
  \begin{split}
   (b &- 2)  R^{(2)4}  - 3 b(b-1) R^{(2)} \left\{ R^{(2)} \left[
   R^{(4)}_{,rr} + \frac{2}{r} R^{(4)}_{,r} \right. \right.
   					\\
   & \left. + \frac{1}{2} R^{(2)}_{,r}
   \left( g^{(2)}_{00,r} + g^{(2)}_{11,r} \right) \right] 
   + 2 (b-2) R^{(2)}_{,r} R^{(4)}_{,r} \bigg\} 
   					\\
   & + 3 b(b-1)(b-2) R^{(2)2}_{,r} R^{(4)} = 0.
  \end{split}
\label{H:2m}
\end{equation}

  The components \(\mathcal{H}^{(2b)}_{\mu\nu}\) of the field
equations~\eqref{einstein:eqs} at order \( \mathcal{O}(2b) \) are given by:

\begin{equation}
  \begin{split}
  &\mathcal{H}_{\mu\nu}^{(2b)} = - b (b -1) \bigg\{ R^{(2)b-2} \bigg[
    R^{(4)}_{,\mu\nu} - \Gamma^{1(0)}_{\mu\nu} R^{(4)}_{,r} -
    \Gamma^{1(2)}_{\mu\nu} R^{(2)}_{,r} 
    					\\
    & -g_{\mu\nu}^{(0)}  \bigg( R^{(2)}_{,r} \left[ g^{11(2)}_{,r}
	+ g^{11(0)} (\ln\sqrt{-g})_{,r}^{(2)} + g^{11(2)}  \right.
					\\
    & \left. \times (\ln\sqrt{-g})_{,r}^{(0)} \right] 
    + g^{11(0)} \left[ (\ln\sqrt{-g})_{,r}^{(0)}R^{(4)}_{,r}
	+ R^{(4)}_{~,rr} \right] 
					\\
    & + g^{11(2)}R^{(2)}_{,rr} \bigg) - g^{(2)}_{\mu\nu} g^{11(0)}
       \bigg( R^{(2)}_{,r} (\ln\sqrt{-g})_{,r}^{(0)} + R^{(2)}_{~,rr}
       \bigg) \bigg]
       					\\
    & + (b-2) R^{(2)b-3} R^{(4)} \bigg[ R^{(2)}_{,\mu\nu} -
      \Gamma^{1(0)}_{\mu\nu} R^{(2)}_{,r} - g^{(0)}_{\mu\nu}  g^{11(0)}
					\\
    & \times \bigg( (\ln\sqrt{-g})^{(0)}_{,r} R^{(2)}_{,r} + R^{(2)}_{,rr}
       \bigg) \bigg] \bigg\} - b(b-1)(b-2) 
      					\\
    & \times \bigg\{ R^{(2)b-3} \bigg[
      2R_{,\mu}^{(2)}R_{,\nu}^{(4)} - g_{\mu\nu}^{(0)} \bigg( 2 g^{11(0)}
      R_{,r}^{(2)} R_{,r}^{(4)} + g^{11(2)} 
     					\\ 
    & \times R_{,r}^{(2) 2} \bigg) - g_{\mu\nu}^{(2)} g^{11(0)} 
      R_{,r}^{(2)2} \bigg] + (b-3) R^{(2)b-4} R^{(4)} 
    					\\
    & \times \bigg[ R^{(2)}_{,\mu} R^{(2)}_{,\nu} -
      g^{(0)}_{\mu\nu} g^{11(0)} R^{(2)2}_{,r} \bigg] \bigg\} .
  \end{split}
\label{Hmu-nu:2m}
\end{equation}

  Dividing the field equations~\eqref{eq7a:notes} by 
\(R^{(2)b-4}\) and using the trace~\eqref{H:2m} and the last
equation, a reduced expression for the field equations is found:

\begin{equation}
\begin{split}
  & - \frac{2b-1}{6} R^{(2)4} + b R^{(2)3} R^{(2)}_{\mu\nu}
    - b (b-1) R^{(2)2} \left[ R^{(4)}_{,\mu\nu} \right. \\
  & \left. - \Gamma^{1(0)}_{\mu\nu} R^{(4)}_{,r}
  - \Gamma^{1(2)}_{\mu\nu} R^{(2)}_{,r} \right] - b (b-1) (b-2) R^{(2)}  \\
  & \times \left[ R^{(4)} \left( R^{(2)}_{,\mu\nu}
  - \Gamma^{1(0)}_{\mu\nu} R^{(2)}_{,r} \right) + 2 R^{(2)}_{,\mu} R^{(4)}_{,\nu} \right]  \\
  & - b (b-1) (b-2) (b-3) R^{(2)}_{,\mu} R^{(2)}_{,\nu} R^{(4)} = 0 \, ,
\end{split}
\label{eq10:notes}
\end{equation}

\noindent which can also be regarded as the traceless component of the
field equations.


\label{lastpage}

\end{document}